# THz probing of non-trivial topological states in Co$_2$MnGe Heusler alloy thin films

**Ekta Yadav, Anand Nivedan, and Sunil Kumar***

*Femtosecond Spectroscopy and Nonlinear Photonics Laboratory,
Department of Physics, Indian Institute of Technology Delhi, New Delhi 110016, INDIA*
*\*Corresponding Author: kumarsunil@physics.iitd.ac.in*

(March 27, 2024)

Co$_2$MnGe (CMG) has been demonstrated recently as a half-metallic ferromagnetic Heusler alloy which possesses a topologically non-trivial band structure. This behavior is unique to such systems and hence warrants extensive experimental exploration for potential spintronic and chirality sensitive optoelectronic applications. Here, we demonstrate that an epitaxial thin film of CMG acts as a source of THz radiation upon photoexcitation by optical femtosecond laser pulses. Detailed experiments have revealed that a large contribution to THz emission occurs due to nonmagnetic or spin-independent origin, however, significant contribution in the THz generation is evidenced through excitation light helicity dependent circular photogalvanic effect (CPGE) confirming the presence of topologically non-trivial carriers in the epitaxial CMG thin films. Furthermore, we show that not only the topological contribution is easily suppressed but also the overall THz generation efficiency is also affected adversely for the epitaxial films grown at high substrate temperatures.

Weyl semimetals, a novel class of quantum materials, hold significant scientific interest due to their unique electronic band structure and intriguing properties.[1,2] When these materials are exposed to optical femtosecond laser pulses, an effective ultrafast photocurrent is induced that can lead to terahertz (THz) radiation generation.[2-7] Detection of the THz pulses with suitable methods can help to understand various intrinsic physical mechanisms undergoing in these materials systems. Even the analysis of the outcomes from these measurements can prove to be beneficial for manipulating the topological material systems and engineering the desired properties out of them for emerging applications.[6,8] Heusler compounds, hosting various non-trivial topological phases can be found to be either topological insulators or semimetals and so on, and can provide high tunability in the electronic properties due to the stability of a broad range of chemical compositions and the stoichiometry.[9,10] Similar to other typical ferromagnetic/nonmagnetic spintronic heterostructures,[11-14] in a few recent reports, an ultrafast photoinduced super-diffusive spin current in the ferromagnetic Heusler layer and the spin-charge conversion in adjacent heavy metal layer have been considered to be responsible for the THz emission from the Heusler alloy/heavy metal bilayer heterostructures.[15,16]

The full-Heusler alloy Co$_2$MnGe (CMG) is a centrosymmetric half-metallic ferromagnet with the Curie temperature above ~900 K [17]. This ternary system of X$_2$YZ stoichiometric composition crystallizes in a cubic structure with space group Fm$\bar{3}$m.[18] Electronic band structure calculations have revealed the presence of Weyl points near the Fermi energy level and these nodes display topologically nontrivial characteristics.[9] Earlier, from angle-resolved photoelectron spectroscopy of bulk CMG, Kono et al.[19] were able to disclose that the system exhibits time-reversal symmetry-broken topological Weyl fermionic states within its electronic band structure. This was well supported from the magnetotransport behavior of the bulk CMG compound in another recent report.[20] The existence of topological non-trivial states in CMG was also supported in our previous demonstration of intrinsically dominated anomalous Hall effect in epitaxial thin films of CMG.[21]

Topological semimetals are known to exhibit excitation light helicity-dependent photocurrent generation in them, a phenomenon resulting from second order nonlinear optical effects.[3] Alternately, the detection of the photocurrents by either electrical or optical means[22,23], one can infer the role of the associated topological charge carriers of the Weyl nodes through the dependence of the photocurrent via the circular photogalvanic effect (CPGE).[4,5,7,24] In fact, polarization sensitive THz emission from CPGE-induced ultrafast photocurrents in optical femtosecond laser excited Weyl semimetals can be studied to distinguish the topological surface states contribution over the majorly bulk contribution to the THz emission.[25,26] For centrosymmetric materials such as CMG, where the bulk second order nonlinear effects are prohibited due to inversion symmetry conservation, THz emission can serve as an effective probe to directly investigate the topological surface states. In the present work, we demonstrate ultrashort THz pulse emission from epitaxial thin films of CMG Heusler alloy grown by optimized pulsed laser deposition technique under different substrate temperatures. We have found that the THz peak amplitudes are closely related to the thin film substrate temperature and hence the chemical-ordering, surface roughness and crystalline quality of the sample. Experiments performed under external static magnetic field have helped to distinguish the spin-ordering dependent and independent contributions to the THz generation. Interestingly, the spin selective THz contribution arising due to the CPGE induced photocurrent gets suppressed nearly completely if the thin films substrate temperature is elevated from 500 °C to 650 °C. For the latter, overall

THz emission is the least, and it can be attributed entirely due to the consequence of the dominating ultrafast demagnetization. The excitation light helicity dependent THz emission response of the films establishes the presence of nontrivial topological surface states in the material system.

Epitaxial CMG films of thickness 30 nm are grown on 0.5 mm thick single side polished MgO (100) substrates by using pulsed laser deposition technique at different substrate temperatures $T_S$ and in a vacuum environment of ~$10^{-6}$ mbar.[21] A CMG target (1 inch diameter and appropriate stoichiometry) having purity 99.9% was ablated using a 248 nm wavelength KrF excimer laser delivering laser pulse shots at 7 Hz repetition rate. A fixed laser energy density of 2.6 J/cm$^2$ was used for the material ablation while the CMG film thickness was controlled by adjusting the number of laser shots on the target. A series of MgO/CMG(30 nm) films were grown at the substrate temperatures of $T_S$ = 500 °C, 600 °C, and 650 °C and correspondingly they are referred to as CMG-500, CMG-600, and CMG-650, respectively, throughout this paper. The crystallographic properties of the CMG films were evaluated by recording the X-ray diffraction (XRD) spectra, while the epitaxial nature was confirmed through φ-scans, both undertaken with the help of a PANalytical X'Pert diffractometer having Cu K$_α$ radiation source at 1.5406 Å wavelength. Figure 1(a) displays the XRD patterns of CMG thin films and bare MgO (100) substrate. The presence of (200) and (400) diffraction peaks at angles 2θ = 31.34° and 65.3°, respectively, indicates the formation of epitaxial films with a B2-type structure oriented along the (100) direction of the underlying MgO substrate. A difference in the intensities of the (200) and (400) diffraction peaks for the films grown at different substrate temperatures can be noticed in Fig. 1(a). The intensity ratio, $I_{200}/I_{400}$ between the (200) and (400) diffraction peaks is ~ 0.4, 0.7 and 0.7 for the CMG-500, CMG-600 and CMG-650 samples, respectively, which is indicative of a weakened chemical-ordering in the B2 phase of the material.[27] We have also found that the average surface roughness of the CMG-600 and CMG-650 films is slightly higher than that of the CMG-500 sample.

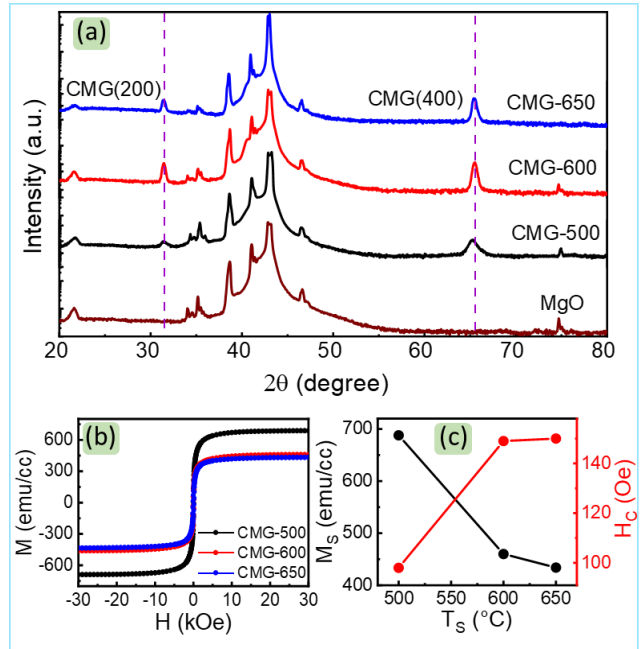

FIG. 1. Crystalline and magnetic properties of CMG-500, CMG-600, and CMG-650 thin films. (a) X-ray diffraction patterns of the different CMG thin films and the MgO substrate. Vertical dashed lines mark the positions of the highly oriented (200) and (400) crystal planes in the CMG lattice. (b) Magnetic hysteresis loops of CMG thin films obtained for in-plan external magnetic field. (c) Saturation magnetization, M$_S$ and coercive field, H$_C$ as a function of the CMG film substrate temperature.

Ferromagnetic hysteresis loops were recorded for all the films from the M-H measurements undertaken in a magnetic property measurement system (Quantum Design Inc.). The results are presented in Fig. 1(b) for an in-plane magnetic field. We may note that the M-H response of different films in Fig. 1(b) is after correcting for the diamagnetic behavior of the MgO substrate. Substrate temperature-dependence of the saturation magnetization (M$_S$) and coercivity (H$_C$) were monitored and the same are presented in Fig. 1(c). It is observed that M$_S$ decreases and H$_C$ increases with the increasing T$_S$. A decrement in M$_S$ and increase in H$_C$ with respect to the increase in the T$_S$ value is in line with the fact that the higher chemical and structural degradation occurs at higher substrate temperatures.[28] Minimum change in chemical-ordering is prone to cause significant modification in the electronic structure of the CMG material system.[9]

THz emission experiments were conducted under ambient conditions of room temperature and air using a home-built time-domain spectrometer[13,14] based on a Ti:sapphire regenerative amplifier delivering 35 fs pulses centered at 800 nm and 1 kHz repetition rate. The main features are captured in the schematic shown in Fig. 2(a). Femtosecond excitation pulse polarization was varied by using a quarter wave plate (QWP) for its fast-axis at angle $\alpha = 0°$, for linear polarized (LP), $\alpha = +45°$ for right circularly polarized (RCP) and $\alpha = -45°$ for left circularly polarized (LCP) states. A constant pump-fluence of ~800 $\mu J/cm^2$ over a collimated beam diameter of 3 mm on the sample was used in all the experiments reported in this paper. In the laboratory XYZ-frame, the THz pulses emitted off the surface of the thin film sample were collected in the reflection geometry by a pair of 90-degree off-axis parabolic mirrors and the associated electric field $E_{THz}(t)$ was captured in real time by electro-optic sampling.[29,30] An in-plane external static magnetic field B = 200 milli-Tesla (2000 Oe), represented by a thick upward arrow along Y-direction in Fig.2(a) was used to magnetize the samples for magnetic field dependent experiments. The orientation of the field could be altered on the plane of the film in a rotatable mount.

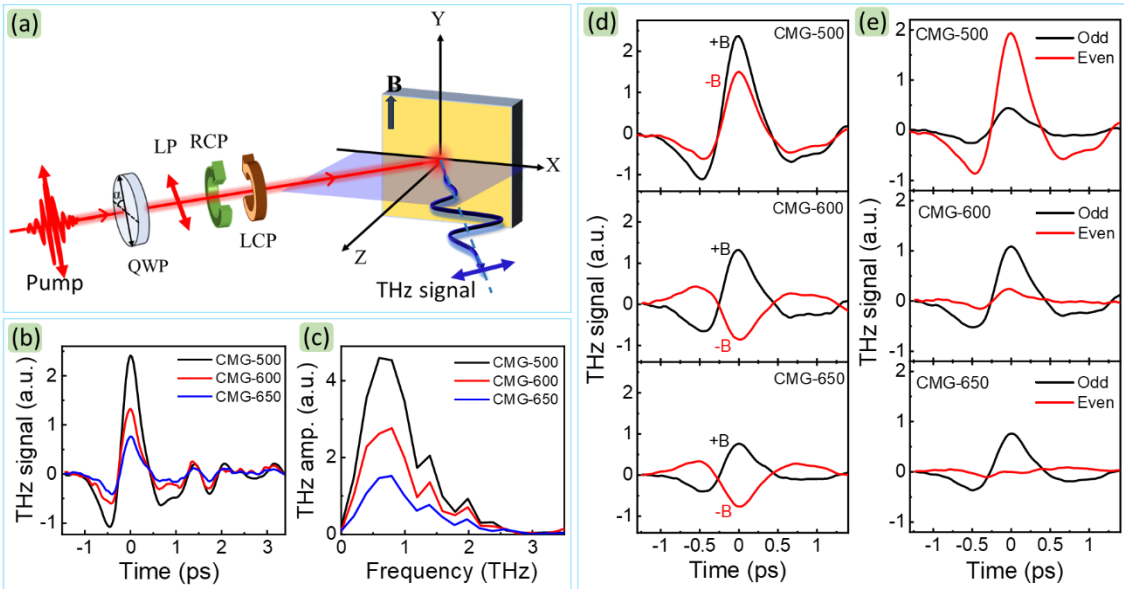

**FIG. 2.** THz emission spectroscopy measurements on CMG-500, CMG-600, and CMG-650 thin films. (a) Schematic diagram of the experimental setup depicting the optical femtosecond laser pulse excitation, THz emission and pick-off in the reflection geometry. Proper rotation (α) of the quarter wave plate QWP ensures different polarization states of the excitation light. (b) Typical time-domain THz signals and (c) the Fourier spectra spanning over above 3 THz bandwidth obtained with the linearly polarized (α = 0°) excitation pulses. (d) THz signals emitted by the photoexcitation films under oppositely directed external static magnetic field ±B = ±2000 Oe. (e) Spin-dependent (odd) and spin-independent (even) components of the THz signals. See text for more details.

Typical time-domain scans $E_{THz}(t)$ and the corresponding angular frequency spectra $E_{THz}(\omega)$ of the THz radiation emitted by CMG-500, CMG-600 and CMG-650 samples for LP excitation pulses are shown in Figs. 2(b,c). A drastic difference in the THz pulse magnitude is seen though the characteristic pulse time-domain shape and the spectral width of ~3THz are mostly identical for all the three samples. This result clearly demonstrates a close relationship between the THz emission efficiency of the films with the substrate temperature Ts and hence the chemical-ordering, the magnetization and surface roughness in them. For the LP excitation case, in Fig. 2(d) we first present and analyze the THz emission response of all the three samples with respect to the B field orientation and later in Fig. 3, we have analyzed the excitation pulse helicity ($\alpha = 0°, +45°, -45°$) dependent response.

THz emission measurements performed under oppositely directed in-plane external magnetic field conditions can help resolve the origin of THz emission from these systems. For +B and -B fields, the THz time-domain scans from the samples are presented in Fig. 2(d). A large difference in the THz signal behaviors for the three films can be immediately noticed. On one hand, there is minimal difference in the THz signal magnitude with respect to the orientation of the B-field for the CMG-500 sample while on the other hand, the THz pulse polarity gets reversed upon inverting the B-field direction for the CMG-600 and CMG-650 samples. Such a behavior of the THz pulse polarity in consequent to the magnetization or the B-field direction in the CMG-600 and CMG-650 is indicative of the strong spin-ordering present in the films.[15,31] Clearly, two intrinsic mechanisms are responsible for the origin of THz generation from the CMG samples. Hence, the spin-ordering dependent (odd) and independent (even) origins to the THz emission are extracted using the following relations, respectively,[2,15]

$$E_{odd}^{THz}(t) = [E_{+B}(t) - E_{-B}(t)]/2,$$
$$E_{even}^{THz}(t) = [E_{+B}(t) + E_{-B}(t)]/2, \quad (1)$$

The extracted values of $E_{odd}^{THz}(t)$ and $E_{even}^{THz}(t)$ are presented in Fig. 2(e) and their corresponding percentage contributions are given in Table I. From this analysis we conclude that the THz emission from CMG-650 sample has mostly entirely a magnetic origin while that from the CMG-500 sample, its primarily due to the nonmagnetic origin.

**TABLE I.** Spin-ordering dependent (odd) and independent (even) contributions to the THz emission from the CMG films. The % contributions, i.e., $E_{odd}^{THz} / (E_{even}^{THz} + E_{odd}^{THz})$ and $E_{even}^{THz} / (E_{even}^{THz} + E_{odd}^{THz})$, are provided inside the brackets.

|  | CMG-500 | CMG-600 | CMG-650 |
|---|---|---|---|
| $E_{odd}^{THz}$ | 0.68 (19.6%) | 1.63 (82.5%) | 1.13 (97.6%) |
| $E_{even}^{THz}$ | 2.80 (80.4%) | 0.35 (17.5%) | 0.03 (2.4%) |

The origin of the spin-independent THz emission from CMG films is attributed to the bulk photon drag effect.[3,32] Such a non-topological effect mediated by third-order non-linear processes has been previously invoked in multiple thin metallic film systems, where an effective photon momentum transfer to the free electrons results into generation of a photocurrent in the direction of the excitation light propagation.[32-34] In addition, the nonzero second order optical susceptibility at the surface can also contribute to the THz emission by optical rectification of the broadband femtosecond excitation pulses, where the magnitude of the effect in this case as well would be dependent on the surface smoothness. Other potential mechanisms for THz emission include photo Dember and drift current effects that are usually known for narrow-gap semiconductors, however, they remain either elusive or largely unexplored in the case of semimetals.[3] For the spin dependent THz emission from the CMG thin film samples, either of or both of the ultrafast demagnetization[35,36] and the photogalvanic effect[3,4,25] can contribute to it. Given that CMG is centrosymmetric in the bulk, hence the photogalvanic effect being a second order nonlinear effect, it can originate from the topological surface states only where the inversion symmetry is broken.[3,25] Indifference in the THz signal magnitude with respect to the excitation light helicity represents the origin majorly due to ultrafast demagnetization.

The excitation light helicity dependent difference in the photocurrents is a direct consequence of circular photogalvanic effect induced photocurrent from the spin polarized topological surface states.[23,26,30,37] To demonstrate that ultrafast CPGE is the source of THz emission from the nontrivial topological states in the CMG films, we now present in Fig. 3 the femtosecond excitation pulse helicity dependent THz emission results for all the three CMG thin films. Figure 3(a) shows the as recorded THz waveforms for excitation light helicities 0, +1 and -1 corresponding to $\alpha = 0°$, $\alpha = 45°$ and $\alpha = -45°$, respectively, where the results for the three samples are shifted along the horizontal axis for a clarity. A large difference with respect to the excitation light helicity is observed. This is attributed to the asymmetric generation of the photocurrents for different helicity light due to depopulation of the carriers from the spin-polarized surface states within the spin-momentum selection rules.[38] The CMG-500 sample shows the largest difference in the THz emission response with respect to the excitation optical pulse helicity. This difference in the THz emission response is found to be relatively weaker for the CMG-600 sample while it's not present at all in the CMG-650 sample. Such an observation makes it quite evident that the CMG-650 lacks the topological surface states while the CMG-500 is a clean system to possess those. In terms of the magnitude of the peak-to-peak amplitude of the THz signals ($E_{pp}$) as indicated in Fig. 3(a), this variations with respect to the CMG film substrate temperature $T_s$ and the excitation pulse helicity are shown explicitly in Fig. 3(b). We may point out that while the observation of the strongest overall THz emission from the CMG-500 sample for the LP case is the same as that presented and discussed above in Fig. 2(b), the THz emission from the CMG-650 sample is the weakest and independent of the excitation pulse helicity.

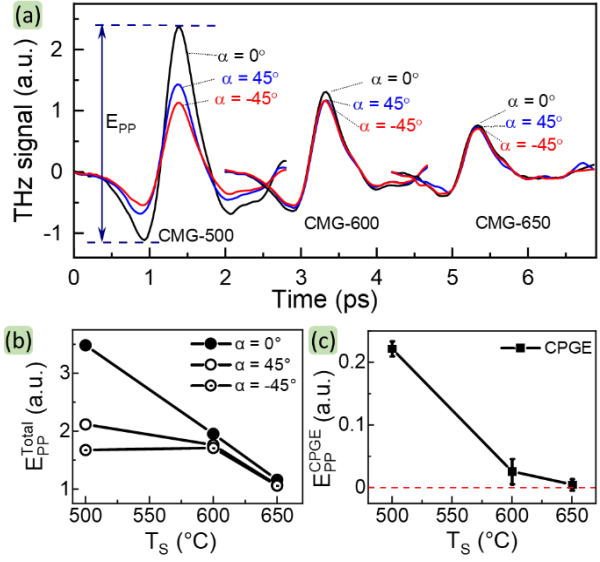

**FIG. 3.** Excitation pulse helicity dependent THz emission from CMG-500, CMG-600, and CMG-650. The in-plane external magnetic field is as shown in Fig 2(a) and only the horizontal component of the THz signals is detected by electro-optic sampling. (a) The experimentally recorded THz waveforms obtained under excitation by LP ($\alpha = 0°$), RCP ($\alpha = 45°$) and LCP ($\alpha = -45°$) light. The data for different samples are shifted on the horizontal axis for better clarity. (b) Variation of the magnitude of the peak-to-peak amplitude of the THz signals ($E_{PP}^{Total}$) as indicated in (a) with respect to the substrate temperature $T_S$. (c) Variation of the CPGE induced contribution to the THz emission with respect to the substrate temperature $T_S$.

Figure 3(c) explicitly shows the relative contribution of the CPGE induced THz emission from the three CMG films extracted using the relation,[29,30] $E_{pp}^{CPGE} = [E_{PP}(\alpha = 45°) - E_{PP}(\alpha = -45°)]/2$, while other bulk contributions related to circular photoexcitation are contained in $E_{pp}^{bulk} = [E_{PP}(\alpha = 45°) + E_{PP}(\alpha = -45°)]/2$. The CMG-500 sample does not only show the highest spin independent THz emission but also demonstrates the highest THz generation efficiency for excitation light of any polarization. The CPGE induced THz emission from CMG-600 and CMG-650 samples is relatively negligibly small. The results suggest that with the increasing sample substrate temperature, the topological state in CMG diminishes progressively and hence CMG-600 and 650 samples have a much-reduced CPGE contribution to the THz emission as compared to that from CMG-500. Interestingly, we find that the relative magnitudes of the $E_{PP}(\alpha = 0°)$, i.e., for the LP excitation, and the $E_{pp}^{bulk}$ are highly correlated for all the three samples under study, which points to the role of the ultrafast demagnetization that can be explored with further experiments. CPGE induced THz emission is well known signature of various topological Weyl semimetals that have been reported in the recent literature.[3,4,7] Therefore, the major assertion from the present ultrafast THz emission studies is that ferromagnetic CMG Heusler alloy possesses nontrivial topological states, and it corroborates very well with our earlier report[21] where we had revealed Berry-phase driven intrinsic anomalous Hall effect in epitaxial CMG thin films.

Helicity dependent photocurrent is also possible by inverse spin orbit torque (ISOT) effect in certain materials in which case the polarization of the generated THz pulse is oriented parallel to the applied magnetic field. Moreover, thus produced THz radiation is typically much smaller than that of the perpendicular component of the THz field emitted from the spintronic heterostructures.[39] In our experimental investigations, the THz detection is set for horizontal polarization of the THz pulses that is perpendicular to the applied magnetic field. Moreover, a photoinduced ISHE has also been invoked in helicity-dependent photocurrent generation, particularly in Bi and related materials.[40] Such an effect cannot be attributed in the present case pertaining to the very short spin diffusion length in Co based Heusler alloys.[41] Therefore, the helicity dependent THz emission by the weak ISOT and photoinduced ISHE mechanisms can be safely ignored..

To summarize, we have studied THz emission properties of epitaxial thin films of CMG ferromagnetic Heusler alloy, where CPGE induced ultrafast photocurrent contribution to THz emission has been explicitly deduced from the excitation light helicity dependent experiments. Particularly, the helicity dependent THz emission from the epitaxial films grown at various growth temperatures of the MgO substrate has evidently shown that the topological surface states can be easily suppressed at high substrate temperatures. These findings will provide valuable insights for developing spintronic and optoelectronic devices based on CMG and similar other ferromagnetic Heusler alloys due to the unique chirality sensitive and topologically protected linear dispersion of Weyl Fermions in them.

**ACKNOWLEDGMENTS**

SK acknowledges the Science and Engineering Research Board (SERB), Department of Science and Technology, Government of India, for financial support through Project No. CRG/2020/000892. Joint Advanced Technology Center, IIT Delhi is acknowledged for support through EMDTERA#5 project. The Physics Department, IIT Delhi is acknowledged for the use of PLD, MPMS and XRD facilities. EY acknowledges the University Grant Commission, Government of India, for Senior Research Fellowship.